# Length Dependence thermal conductivity of Zinc-Selenide (ZnSe) and Zinc Telluride (ZnTe)- A combined first principles and Frequency Domain Thermoreflectance (FDTR) study

Rajmohan Muthaiah[a], Roshan Sameer Annam[a], Fatema Tarannum[a], Ashish Kumar Gupta[b], Jivtesh Garg[a], Ritesh Sachan[b], Shamsul Arafin[c]

a. School of Aerospace and Mechanical Engineering, University of Oklahoma, Norman, OK-73019, USA
b. Department of Mechanical and Aerospace Engineering, Oklahoma State University, Stillwater, OK 74078, USA
c. Department of Electrical and Computer Engineering, Ohio State University, Columbus, OH 43210, USA

**Abstract**:

In this study, we report the length dependence of thermal conductivity ($k$) of *zinc-blende* Zinc-Selenide (ZnSe) and Zinc Telluride (ZnTe) for length scales between 10 nm and 10000 nm using first-principles computations based on density-functional theory. $k$ value of ZnSe is computed to decrease significantly from 11.3 W/mK to 1.75 W/mK as the length scale is diminished from 10 μm to 10 nm. $k$ value of ZnTe is also observed to decrease from 10 W/mK to 1.2 W/mK for the same decrease in length. We also measure the $k$ of bulk ZnSe and ZnTe using Frequency Domain Thermoreflectance (FDTR) technique and observed a good agreement between FDTR measurements and first principles calculations for the bulk ZnSe and ZnTe. Understanding of thermal conductivity reduction at nanometer length scales provides an avenue to incorporate nanostructured ZnSe and ZnTe for thermoelectric applications.

**Introduction**:

With the help of development in improved manufacturing capabilities, the field of microelectronics has advanced significantly[1-3], leading to a significant reduction in the sizes of the transistors being used as predicted by Moore's Law[4]. As the length scale of the materials approach nanometers[5], thermal conductivity exhibited by these materials changes significantly[6-

16].Reduction in size leads to a significant drop in their thermal conductivity[17], leading to an unexpected rise in working temperatures in electronic applications. This phenomenon diminishes reliability and performance microelectronics[18]. Hence, effective thermal management has received significant attention as an area of concern in the advancement of modern microelectronics [7, 9, 12, 15, 19, 20]. Low thermal conductivity can also be beneficial for applications such as thermoelectrics where lower lattice thermal conductivity minimizes heat loss through lattice vibrations, improving the energy conversion efficiency. Many techniques such as Scanning Thermal Microscopy (SThM)[21], Time Domain Thermoreflectance (TDTR)[22-24] and Frequency Domain Thermoreflectance (FDTR)[25] have been used to understand and gain insights into the transport phenomena at nanometer length scales.

Zinc Chalcogenides (ZnX, X=S, Se and Te) are II-VI binary wide band gap semiconductors that crystallize in zinc-blende structure[26-31] and are mainly studied for catalysis[32], electronic[33, 34], structural[34], opto-electronic[35], thermal[29, 36] and thermoelectric properties[37]. In recent times, thin films and nanostructured ZnSe have been widely investigated[28, 37] but their thermal conductivity with size is unknown. In this work, we report the bulk and length dependent lattice thermal conductivity ($k$) of ZnSe and ZnTe by solving phonon Boltzmann transport equation (PBTE) coupled with harmonic and anharmonic interatomic force interactions derived from density-functional theory[38-40]. At 300 K, our first-principles estimated $k$ values of isotopically pure ZnSe and ZnTe are 25.4 Wm$^{-1}$K$^{-1}$ and 14.6 Wm$^{-1}$K$^{-1}$ respectively. The $k$ values of isotopically disordered ZnSe and ZnTe are 23.2 Wm$^{-1}$K$^{-1}$ and 13.7 Wm$^{-1}$K$^{-1}$ respectively. Based on good agreement with the previous work[41], we computed length dependent $k$ of ZnSe and ZnTe by including Casimir scattering[42].

We discuss the bulk and length dependent $k$ of ZnSe and ZnTe using first principles calculations in section A and $k$ of polycrystalline ZnSe and ZnTe using FDTR[25, 43-46] in section B. In Section A, we discuss the phonon dispersion, phonon group velocity, phonon scattering rate and mode dependent contribution of transverse acoustic (TA), longitudinal acoustic (LA) and optical phonons.

For this study, we have also used FDTR to measure the bulk single crystal[25, 45, 46] thermal properties of Zinc Selenide (ZnSe) and Zinc Telluride (ZnTe). FDTR is a non-destructive, non-contact[45] method of measuring the thermal properties of a material. This experimental method utilizes optical pump-probe technique which comprises of a pump laser, which provides the heat to material under study and probe laser which measures the corresponding change in the temperature as function of the changing reflectivity of the material being heated by the pump laser. In FDTR, the pump laser beam is frequency modulated and the measurement is done across a range of frequencies. The frequency of modulation is supplied using a signal generator[43, 46], where in this case, the Lock-In amplifier[43, 46] has the capability of generating input signals of up to 50MHz and the frequency range utilized for this measurement was 2kHz – 50MHz. The sample is coated with a thin layer (~100nm) of gold of high coefficient of thermoreflectance at the probe beam wavelength (532nm). The measurement is done by fitting the measured phase lag, induced by the thermoreflective response of the material φ = φ$_{pump}$ - φ$_{probe}$ to the phase lag that is predicted using the 2D diffusion model, using various input parameters such as transducer thickness, thermal properties of the thin film metal transducer and the effective spot size of the probe beam that was measured during the experiment. Various combinations of thermal and geometrical properties can be extracted with great confidence from this measurement. This measurement is also useful for identifying the isotropy /anisotropy of the materials[47-50]. With its advantages, FDTR is a very versatile measurement technique and can be used to measure the thermal properties of many kinds of material systems such as solids[22, 25, 46, 51, 52], liquids[46, 53, 54], thin films[46, 55-57], and the thermal conductance of interfaces[58-61].

**Computational Details:**

Thermal conductivity in this work is computed by solving the phonon Boltzmann transport equation (PBTE) described below,

$$-\boldsymbol{c}(\boldsymbol{q}s) \cdot \boldsymbol{\nabla} T \left(\frac{\partial n_{qs}}{\partial T}\right) + \frac{\partial n_{qs}}{\partial T}|_{scatt} = 0$$

Where, $n_{qs}$ is the population of phonon mode with wave-vector $\boldsymbol{q}$ and polarization s, $T$ is temperature and $\boldsymbol{c}(\boldsymbol{q}s)$ is the phonon group velocity computed using $\boldsymbol{c}(\boldsymbol{q}s) = \partial \omega_{qs}/\partial \boldsymbol{q}$, ($\omega_{qs}$ represents the frequency of phonon mode $\boldsymbol{q}s$). The equation represents a balance between change in phonon population due to temperature gradient (first term in the equation) and change in population due to scattering (second term). In this work, PBTE is solved exactly for phonon population using QUANTUM ESPRESSO thermal2 code. Knowledge of perturbed phonon populations allows computation of thermal conductivity.

The only ingredients required to compute all phonon properties involved in PBTE are the second order (harmonic) and third-order (anharmonic) interatomic force interactions. These force interactions were derived accurately in this work from first-principles. Harmonic force interactions are needed to compute phonon frequencies, group velocities and populations while anharmonic force interactions enable computation of phonon scattering. Harmonic interactions were derived using the PHONON code, while anharmonic interactions were derived using D3Q package, both within the QUANTUM ESPRESSO density-functional theory (DFT) code.

All the first principles calculations were performed using plane-wave based QUANTUM ESPRESSO[62] package. We used local density approximation (LDA)[63] and norm-conserving pseudopotentials for electronic calculations. Calculations were carried out with plane-wave energy cut-off of 80 Ry and Monkhorst-Pack[64] $k$-point mesh of 12 x 12 x 12 was used for integration over Brillouin zone. Crystal structure was relaxed until the total force acting on each atom diminished below 1e-5 eV/Å. Optimized lattice constants of *Zinc-blende* ZnSe and ZnTe were determined to be a=5.6 Å and a=6.138 Å. Dynamical matrices were computed on 8 x 8 x 8 **q**-grid mesh. Inverse Fourier transform was used to convert these matrices in **q**-space into $2^{nd}$ order force constants (harmonic) in real space and on an 8 x 8 x 8 supercell. Anharmonic ($3^{rd}$ order) interatomic force constants were similarly first computed on 4 x 4 x 4 **q**-grid using QUANTUM ESPRESSO D3Q[65-67] package and then inverse-Fourier transformed to obtain the force constants in real space. Second order (harmonic) force constants are used to compute phonon properties such as phonon frequencies, phonon group velocities and phonon populations. Lattice thermal conductivity is calculated by solving Peierls-Boltzmann transport equation (PBTE)[65, 67, 68]iteratively within QUANTUM ESPRESSO thermal2 code[65-67] with 30 x 30 x 30 **q**-mesh. Casimir scattering[42] is imposed for length dependence $k$ calculations.

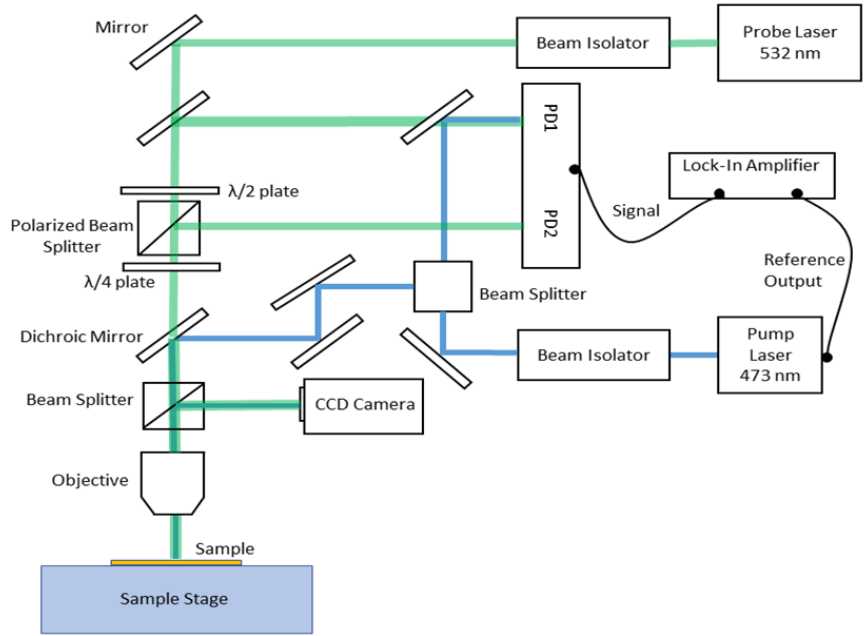

**Figure 1.** Frequency Domain Thermoreflectance (FDTR) experimental setup

**Experimental Setup[45, 46]:**

The FDTR system setup consists of two 20mW continuous wave free space diode lasers (Coherent OBIS) where the pump laser works at the wavelength of 473 nm and the probe laser works at a wavelength of 532 nm. The operating wavelength represents the wavelength associated with visible spectrum range and hence the pump laser at 473 nm is a blue light laser and the 532 nm probe laser is green in color. The pump laser is modulated digitally with the help of output reference of the Lock-In amplifier (Zurich Instruments HF2LI). The driving signal used for pump laser modulation is a sinusoidal signal with a 2V peak to peak voltage and the frequencies used for the modulation range from 2 kHz to 50 MHz. The beams from these lasers pass through optical isolators (Thorlabs IO-5-532-HP for 532nm, Thorlabs IO-3-780-HP for 785 nm), which prevent any backscattering of laser beam into the aperture of the laser, which in turn protects the laser from output power instabilities. The setup also utilizes mirrors to appropriately direct the laser beam in the desired direction. Using a beam splitter, 1% of the pump laser beam is directed towards a photodetector which records the phase of the pump beam, referred to as pump phase or $\varphi_{pump}$. A dichroic mirror (Edmund Optics, hot mirror) reflects the pump beam onto the sample through the microscopic objective. This creates a periodic heat flux on the sample's surface with spot intensity being maximum in the middle and decaying towards the edges according to a Gaussian spatial distribution. Both the pump and probe beams are aligned coaxially, and the probe beam is used to measure the change in phase of the temperature response of the surface with respect to the phase of the input heat signal. This is achieved by measuring the surface temperature through a measurement of the intensity of the reflected probe beam (the two are correlated through the fact that surface reflectivity is a function of surface temperature). Every sample whose thermal properties are measured using FDTR utilizes a thin film of metal, whose thickness is in the range of 80-100 nm. The metal film maximizes the coefficient of thermo-reflectance at the probe's operating wavelength. Balanced photodetection is implemented to improve the signal to noise ratios at the low frequencies of modulation for the pump laser. Balanced photodetector like

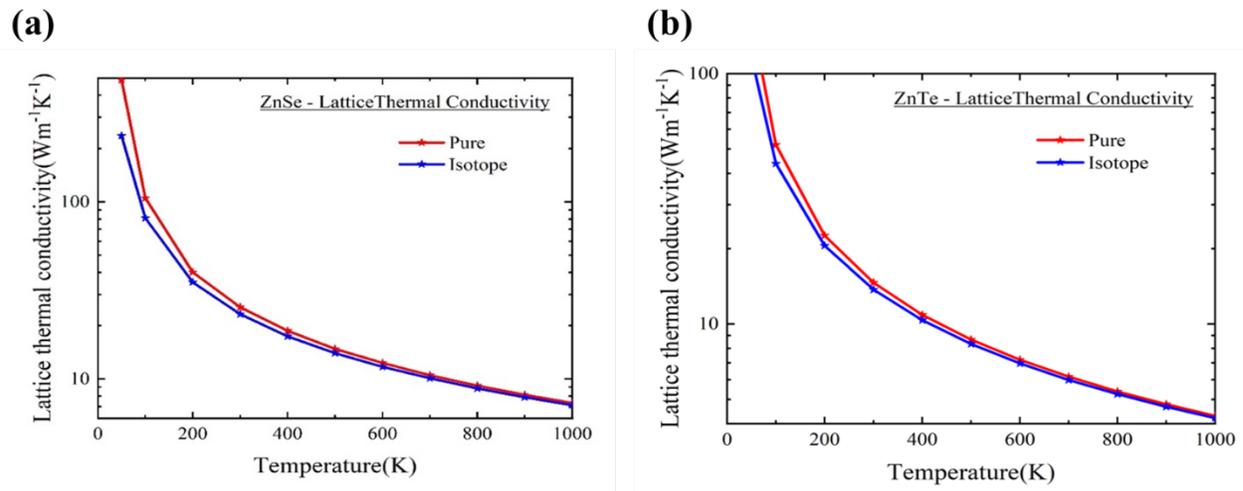

**Figure 2.** Lattice thermal conductivity of pure and naturally occurring: a) ZnSe and b) ZnTe

Thorlabs PDB415A is used, as it consists of two well matched photodiodes namely, PD1 and PD 2. A polarizing beam splitter (PBS) is utilized to split the probe beam into a pre-sample beam and a post-sample beam. The pre-sample beam is reflected into PD 2 as shown in the schematic (Fig. 1). The post-sample beam is reflected from the sample surface, and reflected again using the dichroic mirror into PD 1. In order to reduce noise in the PD signal, the optical path lengths to PD 1 and PD 2 need to be matched. The noise minimization is achieved by delicately balancing the pre and post sample beams, by adjusting the half-wave plate. Another way of noise rejection that has been implemented is by letting the PD 1 and PD 2 subtracted signal pass through a low noise transimpedance amplifier. This method helps eliminate noise from the probe beam, which is a common problem. To prevent the thermal signal from getting overwhelmed by the backscattering of the pump beam, two band pass filters (FGB37) are used before the photodetectors.

The measurement of FDTR in our case works by comparing the phase lag of probe beam's post-sample beam (as discussed earlier) with respect to the phase of the periodic pump heat source. It is important to note that all the components within the experimental setup along with the path lengths traversed by both the pump and probe beams introduce a frequency-dependent phase shift to the signal which is denoted by $\varphi_{ext}$. The way this phase shift is accounted for is by sampling 1% of the pump beam and directing it into the post-sample photodetector. $\varphi_{ext}$ is then measured over a range of the modulation frequencies for the pump beam and is then subtracted from the phase signal that has been measured, before fitting the measured data to a thermal model. In the setup, we also use a translation stage to adjust the sample height which keeps the sample in sharp focus i.e., the sample is in the depth of focus of the objective lens being used. This step is of utmost importance as a sharp image of a focused sample helps in the accurate determination of the spot size, which is a vital input parameter for the thermal model. The image of the sample being focused is done by using the CCD (charge-coupled device) camera.

The ZnSe and ZnTe bulk crystals were purchased from MTI Corporation. The gold pellet for the deposition of gold[45, 48, 69] thin film on bulk polycrystalline ZnSe was purchased from Kurt. JLesker Company (99% purity). We use Lesker Nano36 Evaporator thermal evaporator, (part of Microfabrication Research & Education Center at University of Oklahoma) to deposit 100nm of gold film. After this, the sample was placed carefully on a glass slide using double sided carbon tape, making it ready for measurement using FDTR.

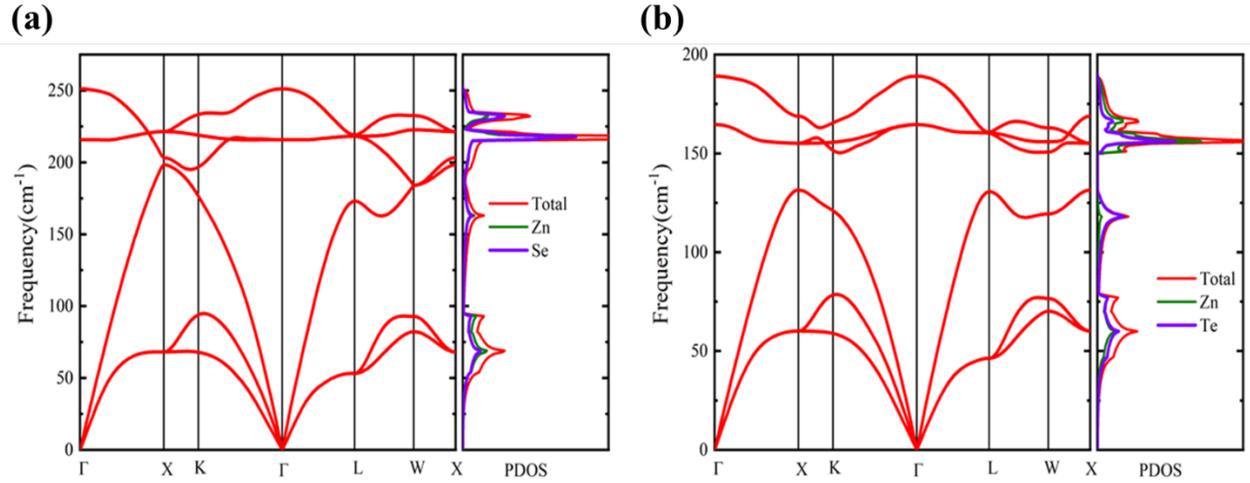

**Figure 3.** Phonon Density of States of a) ZnSe b) ZnTe

**Results and Discussions:**

Temperature dependent lattice thermal conductivity($k$) of Zinc Selenide (ZnSe) and Zinc Telluride (ZnTe) derived from first-principles computations is shown in Figs. 2a and Fig 2b respectively. At room temperature (300K), $k$ of 23.2 Wm$^{-1}$K$^{-1}$ is computed for the naturally occurring ZnSe in close agreement with the previously reported experimental value [29]. Length dependent $k$ of ZnSe nanostructures was computed from first principles calculations by imposing Casimir/boundary scattering[42] and is shown in Fig 4a. At 100 nm, $k$ of 7.04 Wm$^{-1}$K$^{-1}$ is quite comparable to the conventional thermal interface materials[70]. $k$ value of ZnSe is observed to decrease significantly from 11.3 W/mK to 1.75 W/mK as the length scale is diminished from 10 μm to 10 nm (Fig. 4a). Bulk $k$ of ZnTe is computed to be 13.72 Wm$^{-1}$K$^{-1}$ for the naturally occurring case and is again in close agreement with the previously reported values[41]. $k$ value of ZnTe is also observed to decrease significantly from 10 W/mK to 1.2 W/mK as the length scale is diminished from 10 μm to 10 nm (Fig. 4b). We investigated the phonon dispersion, phonon group velocity, phonon scattering rate and phonon mean free path in detail to explain these thermal conductivity results.

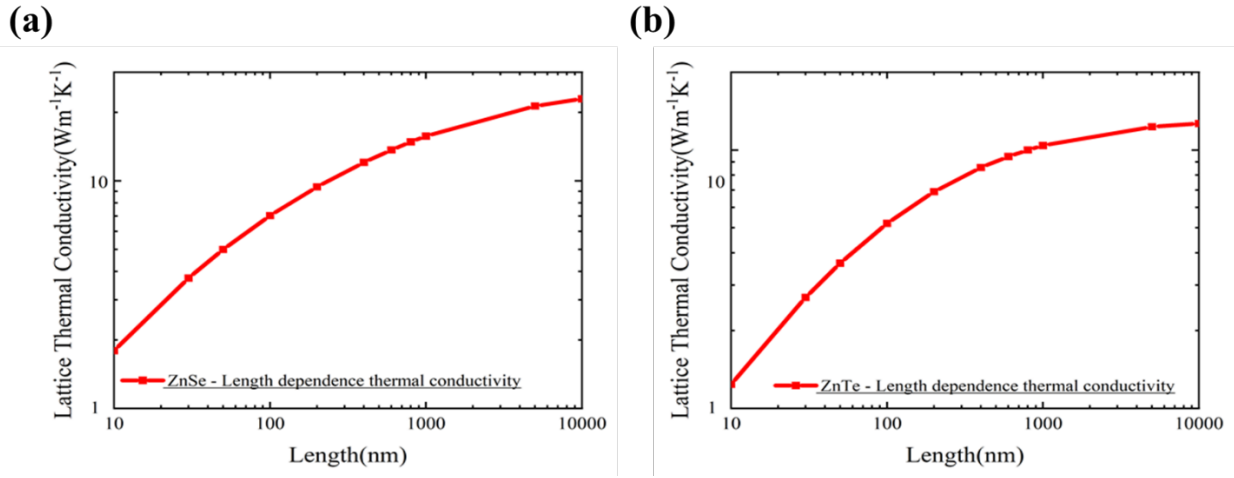

**Figure 4.** Length dependent thermal conductivity of isotopically disordered a) ZnSe b) ZnTe

Contributions of transverse acoustic (TA), longitudinal acoustic (LA) and optical phonons to overall thermal conductivity are shown in Figs. 5a and b for ZnSe and ZnTe respectively.

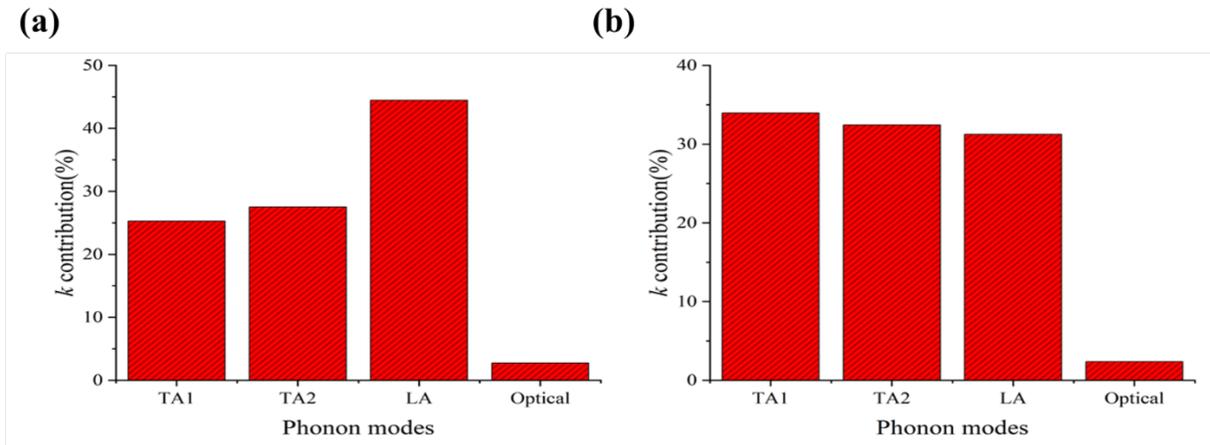

**Figure 5.** Thermal conductivity contribution from TA, LA and optical; phonon modes for a) ZnSe and b) ZnTe

Phonon line width (inverse of lifetime) are shown in Fig 6a and 6b and phonon group velocities are shown in Fig 7a and 7b for ZnSe and ZnTe respectively. In ZnSe, the contribution of LA phonons to overall thermal conductivity is higher than TA and optical phonons (Fig. 5a). This can be understood by noticing that LA phonons have higher phonon group velocities (Fig. 7a), and their scattering rates (Fig. 6a) are lower than TA phonons (in the low frequency range).

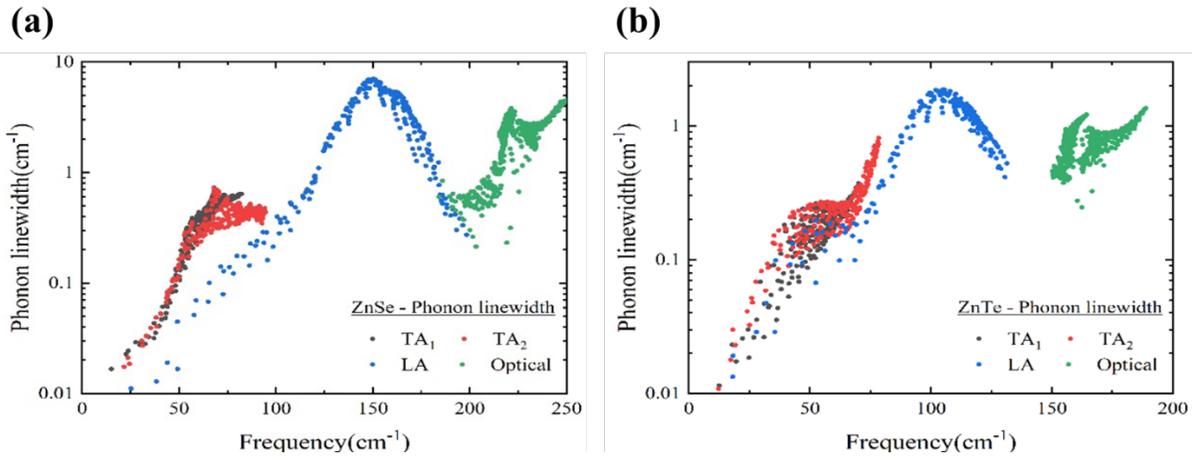

**Figure 6.** Phonon scattering rates of a) ZnSe b) ZnTe

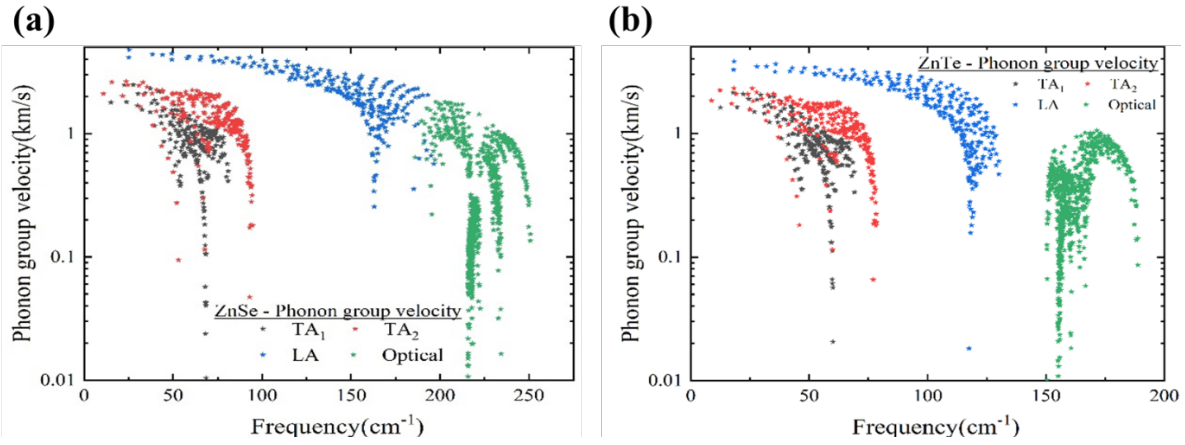

**Figure 7:** Phonon group velocities of a) ZnSe b) ZnTe

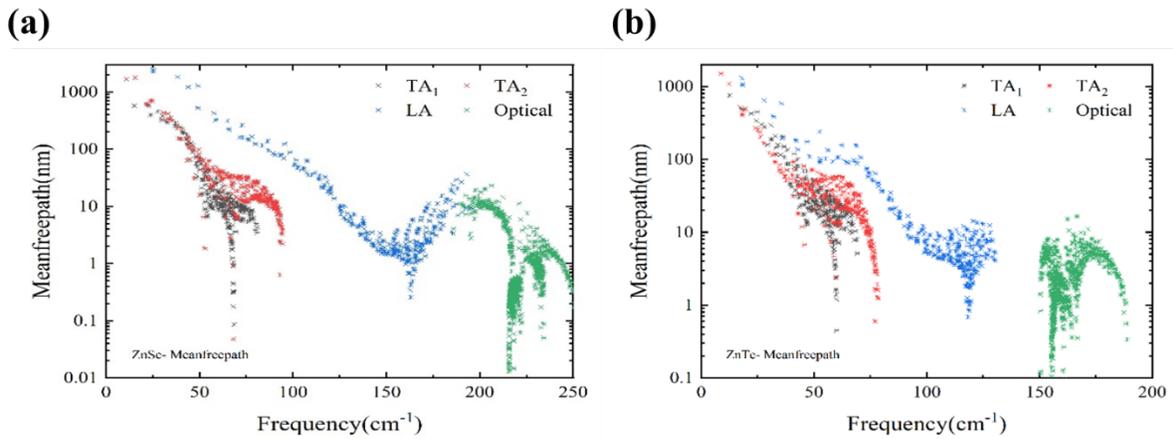

**Figure 8.** Phonon Mean free paths of a) ZnSe b) ZnTe

LA phonons contribute almost 44.5% to overall thermal conductivity in ZnSe. In case of ZnTe, the contribution of TA1, TA2 and LA phonons are ~34%, 32% and 31% respectively at 300 K.

**FDTR Measurement**: The experimentally measured thermal conductivity values of Zinc Selenide (ZnSe) and Zinc Telluride (ZnTe) using FDTR were ~17 Wm$^{-1}$K$^{-1}$ and 14 Wm$^{-1}$K$^{-1}$ respectively at room temperature. The measurement of thermal conductivities of polycrystalline ZnSe and ZnTe using FDTR is achieved by curve fitting method as explained in the experimental setup. The blue circles (in Fig. 9) represent the phase-lag data that has been measured using the FDTR setup where in the frequency of lock-in amplifier was set at 1MHz and the laser power of the pump and theprobe lasers were set at 20mW and 5mW respectively for all samples measured in this work. The solid line represents the best fit curve, which is obtained by a 2D heat conduction mathematical model which utilizes known input parameters such as volumetric heat capacity of the gold film, gold film thermal conductivity, film thickness, volumetric heat capacity of the material under investigation, the thickness of the material under investigation and the laser spot size. The properties of gold film and its thickness are determined by depositing the gold film over a substrate with known properties, namely, fused Silica ($k$ ~1 W/mK) and Sapphire (k ~35 W/mK).

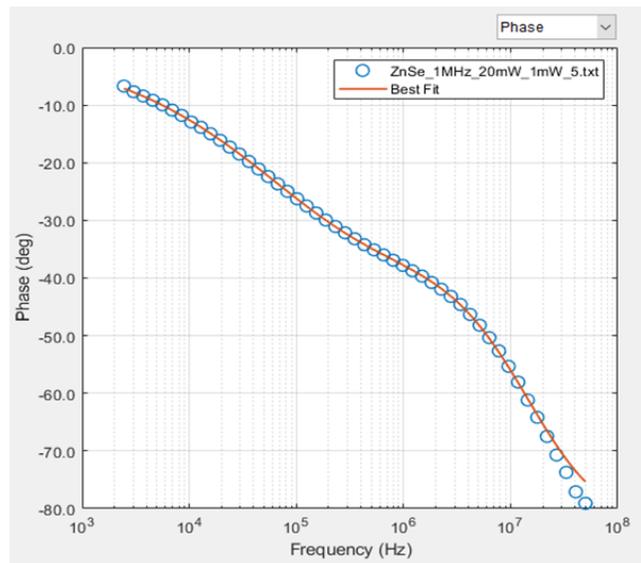

**Figure 9.** FDTR data for curve fitting analysis to calculate the thermal properties of ZnSe measured at a lock-in frequency of 1MHz along with pump and probe powers at 20 and 1mW respectively

This results in accurate determination of the properties of the gold thin film. The volumetric heat capacities of ZnSe and ZnTe are obtained by multiplying the density (g/cm$^3$) and specific heat (J/Kg K). These details are available on the datasheet of the ZnSe and ZnTe on MTI Corporation's website [66]. The only unknown property in our study is the thermal conductivity of ZnSe and ZnTe, which leads to a single parameter fit.

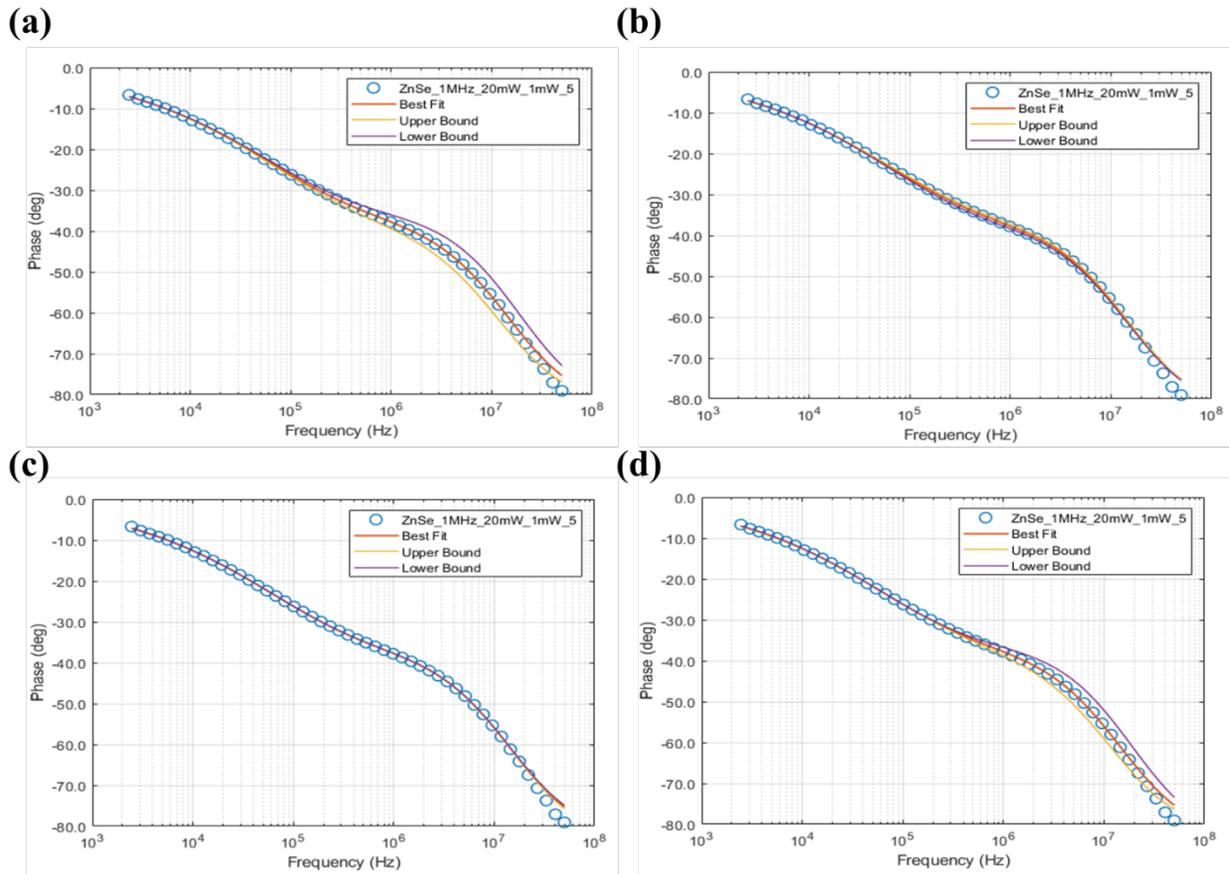

**Figure 10.** Sensitivity of phase-lag to a) gold volumetric heat capacity b) gold cross plane thermal conductivity c) gold in-plane thermal conductivity and d) thickness of gold film deposited on top of the ZnSe crystal.

For the measured thermal conductivity to be accurate, the measured data and the predicted phase-lag from heat conduction model should agree very well. As shown in Figure 9, the best fit curve agrees very well with the measured phase-lag values (represented by blue circles). The figure is representative of the agreement between the measured phase data and the phase data generated using the mathematical model which utilizes the input parameters mentioned earlier in the section.

We next describe the sensitivity analysis which is important in ascertaining the accuracy of measured thermal conductivity values. Insensitive parameters can be fixed as input parameters in the model as they do not affect the curve fitting, and this reduces the number of free parameters to be fitted. Sensitive parameters are seen to have a significant difference in the upper and lower bound values of the phase lag measured in degrees (Fig. 10a, 10d), while insensitive parameters demonstrate small difference between these bounds (Fig. 10b, 10c). Figures 10a, 10b and 10d show us that the measurement is sensitive to gold thin film's volumetric heat capacity, gold's cross-plane thermal conductivity and its thickness (essential input parameters). Figures 11a and 11b, show that the measurement is sensitive to the cross and in-plane thermal conductivity of bulk polycrystalline ZnSe. Figures 12a and b represent another way of understanding sensitivity in terms of the phase difference values.

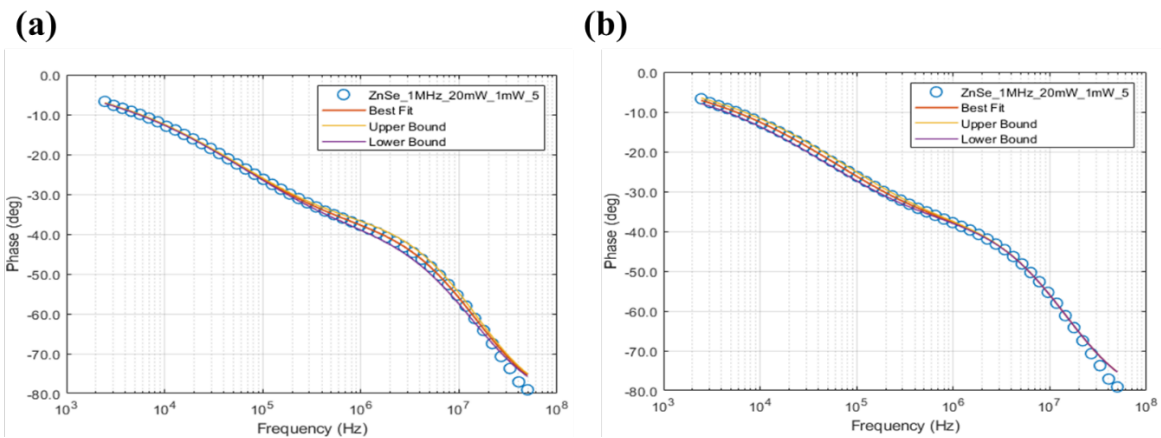

**Figure 11.** Sensitivity of a) Cross Plane b) In Plane thermal conductivity parameters of ZnSe Crystal

As seen in figure 12, the sensitivity analysis shown in figures 10 and 11 can be represented as a function of the phase lag set to a certain upper and lower bound limit, which in this case is set at 20%. This representation is a great way of determining the sensitivity of FDTR to the material parameters. Comparing the figures 10b and 10c with 12a, the orange and yellow curves represent the sensitivity of FDTR to the in-plane and cross plane thermal conductivity of gold thin film.

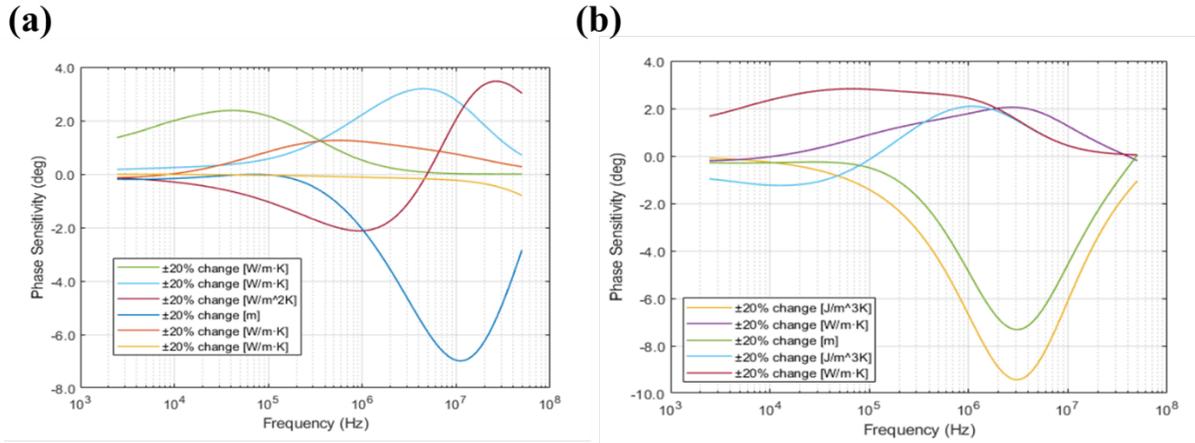

**Figure 12.** Sensitivity analysis of all the parameters of gold transducer and a) ZnSe and b) ZnTe as a function of phase change with limits of ±20% upper and lower bounds

The green and light blue curves in figure 12 corresponds to the FDTR's sensitivity to the in-plane and the cross plane thermal conductivity of ZnSe.

Analysis process similar to that of ZnSe was conducted for ZnTe as well in order to determine its thermal conductivity. Figure 13 shows that with the help of accurately determined input parameters, we were able to achieve a good agreement between the FDTR's measurement of the phase lag and the phase lag calculated using the 2D diffusion model, keeping the thermal conductivity of ZnTe as a floating or a free parameter. This gives us a high confidence value of the thermal conductivity of ZnTe which was measured to be ~14 W/mK.

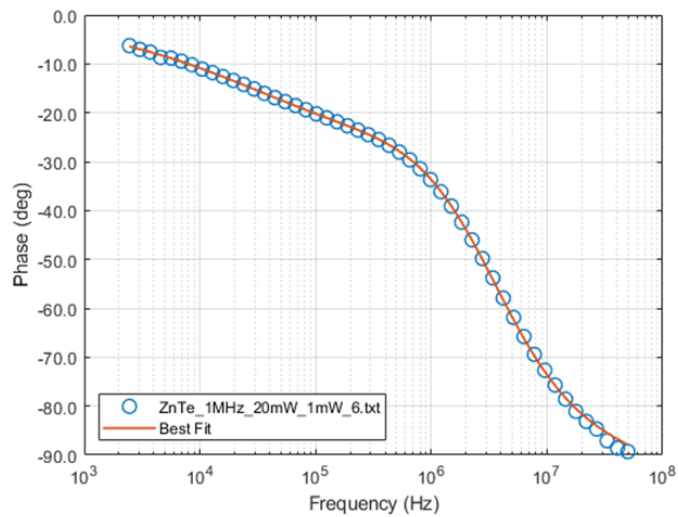

**Figure 13.** FDTR data for curve fitting analysis to calculate the thermal properties for ZnTe

**Conclusion:**

Using FDTR, the experimentally determined thermal conductivity values of ZnSe and ZnTe are found to be ~17 $Wm^{-1}K^{-1}$ and 14 $Wm^{-1}K^{-1}$ respectively. These compare well to the computed results of around 23.2 $Wm^{-1}K^{-1}$ and 13.72 $Wm^{-1}K^{-1}$ for ZnSe and ZnTe respectively at 300K achieved using first-principles. The length dependent k value of ZnSe saw a significant reduction from 11.3 W/mK to 1.75 W/mK with the length scale going from from 10 μm to 10 nm whereas for ZnTe, the *k* value was observed to decrease significantly from 10 W/mK to 1.2 W/mK with similar length scale reduction as ZnSe. The difference in the experimental and the computational results can be attributed to the dissimilarity in the polycrystallinity as obtained from the manufacturer and as modeled for computational analysis. This work is also a definite demonstration of the power tool that is FDTR, which can very precisely and accurately measure the thermal properties of materials. FDTR can also be utilized to measure the thermal properties of some unknown material systems with great confidence and supplemented by the computational work, one can easily gain a greater understanding towards the physics behind the heat transport phenomena taking place within the material system under study. Additionally, this combination of computational work and the experimental work using FDTR can greatly influence the ability of tuning the materials to achieve desired results for specific applications in the areas of thermal management and thermoelectric devices.

**Conflicts of Interest**

There are no conflicts of interest to declare.

**Acknowledgements**

RM, RSA, FT and JG acknowledge support from National Science Foundation award under Award No. #2115067. We also acknowledge OU Supercomputing Center for Education and Research (OSCER) for providing computing resources for this work.